\documentclass{elsart}
\usepackage{natbib}
\usepackage[dvips]{graphicx}
\usepackage{amssymb}

\begin{document}
\begin{frontmatter}
\title{Dark matter search experiment with CaF$_2$(Eu) scintillator at Kamioka Observatory}
\author[Physics]{Y. Shimizu\corauthref{cor1}},
\ead{pikachu@icepp.s.u-tokyo.ac.jp}
\author[Physics,RESCEU]{M. Minowa},
\author[Physics]{W. Suganuma}\footnote{Now at Third Research Center, Technical research and Development Institute, Japan Defence Agency, 1-2-10, Sakae, Tachikawa, Tokyo 190-8533, Japan},
\author[ICEPP]{Y. Inoue}
\address[Physics]{Department of Physics, School of Science, University
of Tokyo, 7-3-1, Hongo,Bunkyo-ku, Tokyo 113-0033, Japan}
\address[RESCEU]{Research Center for the Early Universe(RESCEU), 
School of Science,
University of Tokyo, 7-3-1, Hongo, Bunkyo-ku, Tokyo 113-0033, Japan}
\address[ICEPP]{International Center for Elementary Particle Physics(ICEPP), University of Tokyo, 7-3-1 Hongo, Bunkyo-ku, Tokyo 113-0033, Japan}
\corauth[cor1]{Corresponding author.}
\begin{abstract}
We report recent results of a WIMP dark matter search experiment using 
310g of CaF$_2$(Eu)
scintillator at Kamioka Observatory. 
We chose a highly radio-pure crystal, PMTs and radiation shields, so that
the background rate decreased considerably. 
We derived limits on the spin dependent WIMP-proton and WIMP-neutron coupling 
coefficients, $a_{\rm p}$ and $a_{\rm n}$. 
The limits excluded a part of the parameter space allowed by the annual 
modulation observation of the DAMA NaI experiment. 
\end{abstract}
\end{frontmatter}

\clearpage

\section{Introduction}
There is substantial evidence that most of the matter in our Galaxy must be 
dark matter that exists in the form of Weakly Interacting Massive Particles
(WIMPs) [1]. WIMPs are thought to be non-baryonic particles, 
and the most 
plausible candidates for them are the lightest supersymmetric particles.
WIMPs can be directly detected through elastic scattering with nuclei in 
radiation detectors. Using various detectors, many groups have performed 
experiments for the detection of WIMPs. 

For the direct WIMP detection, two kinds of interactions need to be 
considered: the axial-vector (Spin-Dependent, SD) interaction and the scalar
(Spin-Independent, SI) interaction. WIMPs couple to the spin of the target 
nuclei 
in the SD interaction while they coherently couple to almost all nucleons 
of the target nuclei in the SI interaction. 
For the SD interaction, $^{19}$F is one of the most favorable nuclei to 
detect WIMPs because of its large nuclear spin and 100\% natural abundance. 
Furthermore, the spin 
expectation values of protons and neutrons in $^{19}$F have the opposite signs
while those values in other nuclei usually have the same signs 
[2--6]. Therefore, 
experiments with a $^{19}$F-target can set complementary limits to those 
with nuclei whose spins have the same sign.

Several direct WIMP searches using $^{19}$F-based detectors, such as 
bolometers [7,8], scintillators [9,10] and 
superheated droplet detectors (SDDs) [11,12], have already been 
performed. Our group had also carried out WIMP search 
experiments using LiF and NaF bolometers [7,8]. 
Although these results set complementary limits to those
of NaI(Tl) experiments, the background rates below 20 keV could not be 
decreased sufficiently. The sources of the main background were thought not to 
be any radiation but to be intrinsic and instrumental noises in the 
bolometers. As an alternative method,
we have carried out experiments using CaF$_2$(Eu) scintillators. 
CaF$_2$(Eu) is 
known as a useful scintillator for WIMP searches because of its high light 
yield (19000 photons/MeV) and has already been used in 
several experiments. However, these experiments show rather high background 
rates ($\gtrsim$10 counts/k.e.e./day/kg \footnote{In this letter, we use 
the unit k.e.e. or keV electron equivalent for nuclear recoil energy
since the scintillation efficiency for nuclear 
recoils is less than electron recoils due to high ionization loss.}) 
in comparison with 
the NaI(Tl) experiments ($\sim$ 1 counts/k.e.e./day/kg).
In addition, its light yield is about 50 \% of NaI(Tl) , so that the energy 
threshold of CaF$_2$(Eu) detectors is higher than NaI(Tl) detectors. 
Therefore, these CaF$_2$(Eu) experiments set less 
stringent limits to the SD interaction than the NaI(Tl) experiments.
In order to reduce the background rates and to improve the energy threshold of
our experiments, we have tried to eliminate
radioactive sources of $\gamma$-rays and studied
detection efficiency near the energy threshold.

In this letter, the results of our new experiment with a CaF$_2$(Eu) 
scintillator are presented.

\section{Experimental setup}
\begin{figure}[thb]
\begin{center}
\includegraphics[width=14cm]{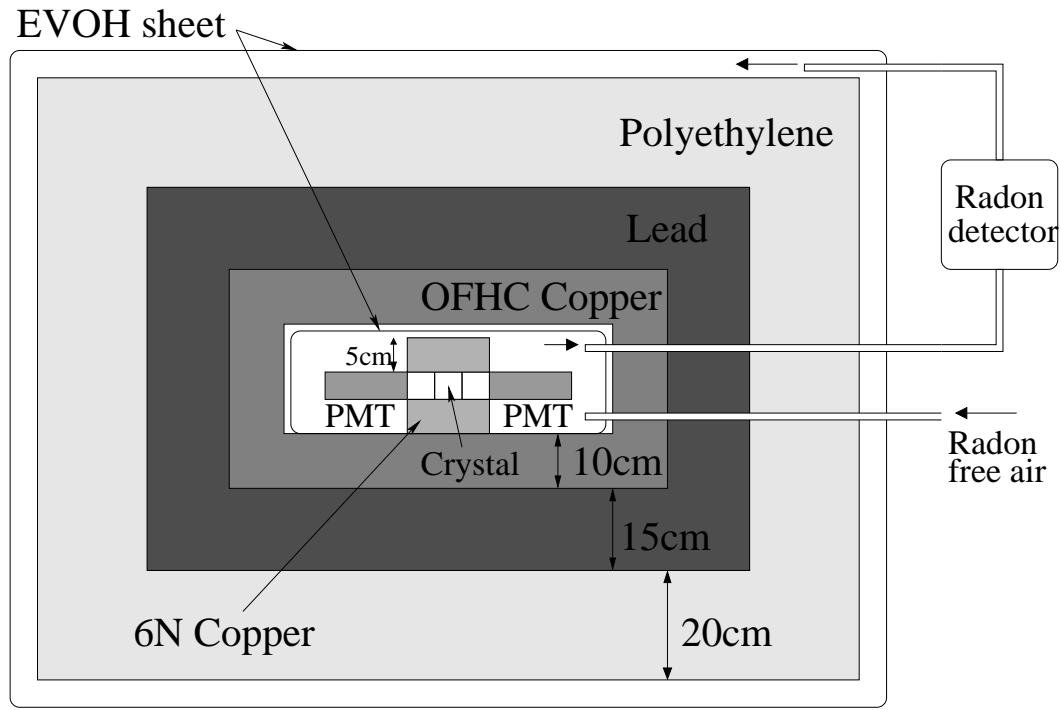}
\caption{Schematic view of the experimental setup.}
\label{setup}
\end{center}
\end{figure}

The detector consisted of a CaF$_2$(Eu) crystal with a mass of 310g.  It was
installed in Kamioka Observatory (2700 m.w.e.). To reduce the background,
low radioactive PMTs (Hamamatsu R8778) and a low radioactive radiation 
shield were used. The experimental setup 
is shown in Fig \ref{setup}. 
The crystal was produced from the same stock of CaF$_2$ raw powder as  
used by the CANDLES experiment [14] and low radioactive EuF$_3$ 
powder. The low 
radioactive PMT was developed by the XMASS experiment [13] and 
thought to be the 
least radioactive among available PMTs. Two PMTs were attached to the crystal
through 5cm-long quartz light guides. The photoelectron yield of the 
scintillator was about 4 photoelectrons/k.e.e. at 60 k.e.e.
The radiation shield consisted of 5cm of 
highly pure copper (99.9999\%), 10 cm of OFHC copper, 15 cm of lead, and 
20 cm of polyethylene. The highly pure copper was supplied by Mitsubishi
Materials.
The whole setup was separated from the mine air by two
layers of EVOH sheet [15] so that radon gas in the air could not come 
into the detector. The radon free air
generated in Kamioka observatory was sent into the detector as shown in the 
figure. In addition, the radon concentration around the detector was 
continuously monitored by a radon detector. It was kept at about
30 mBq/m$^3$ throughout the experiment. 

For the data acquisition, a dual trace digital oscilloscope was used 
to record wave forms 
of PMT pulses. 
The pulses of each PMT are directly sent to the oscilloscope without any
amplifier then daisy-chained to a trigger system with a tee connector.
The impedance of each oscilloscope input was set to 1M$\Omega$ and the 
cables are terminated at the inputs of the trigger system. The wave forms
were digitized at the oscilloscope at a rate of 1 GS/s for a total 
digitization time of 10 $\mu$s and sent to a PC for an off-line analysis. 

In the trigger system, the pulses were amplified by
PMT amplifiers and sent to low threshold discriminators.
Thresholds of the discriminators were set to 1/4 of the mean pulse hight
of single photoelectron events. 
The single photoelectron events were obtained by irradiating the PMT with
continuous feeble light from an LED.
The detection efficiency of each PMT with the discriminator threshold
for a single photoelectron event was obtained to be 0.8 by comparing 
the discriminator output counts and the total number of single photoelectron
events; the latter was estimated by assuming a normal distribution for the
charge distribution of the single photoelectron events. Parameters of 
the normal distributions were determined in a special measurements 
by recording the events with low-threshold internal triggers of the 
oscilloscope to include the low energy tail of the spectrum.
The charge distribution of the internally triggered events was well
fitted by the normal distribution. 
Single counting rates of PMTs were typically 20 Hz.

The discriminator signals were then sent to the coincidence circuit.
It generated a trigger pulse for the oscilloscope when both of the PMTs
gave signals in coincidence within 1$\mu$s.
It reduced the number of background events due to dark noise of the PMTs.

The trigger efficiency of the data acquisition system was calculated as a
function of a number of photoelectrons by the Monte-Carlo analysis.
In this analysis, we generated a certain number of photoelectrons 
in a time sequence according to the 
scintillation decay time of CaF$_2$(Eu). We assumed each photoelectron 
reached in each PMT evenly. 
The detection probability of a photoelectron was set to 0.8.
Then, we measured an interval between arbitrary photoelectrons of each PMT
in the event.  
The fraction of events that generated at least one output signal in both of
the discriminators within 1 $\mu$s was taken to be the trigger efficiency.
At 2 k.e.e., 8 photoelectrons were generated and the trigger efficiency 
was calculated to be 0.93. 
Threshold energy for the following analysis was set at 2 k.e.e.

\section{Measured spectra}
\begin{figure}[thb]
\begin{center}
\includegraphics[width=14cm]{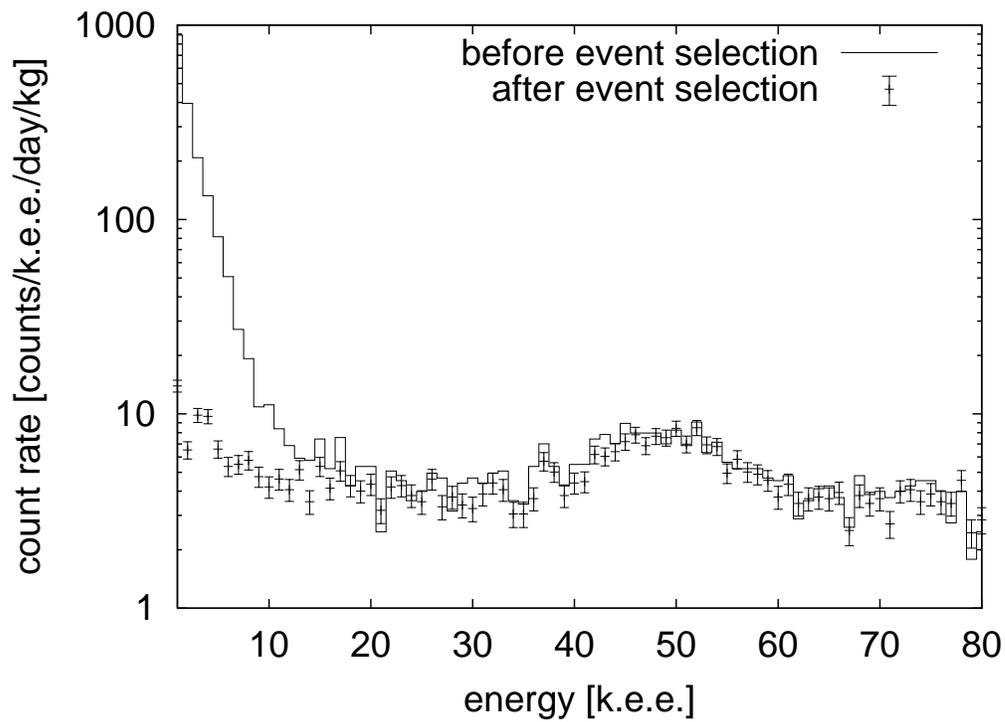}
\label{spectra}
\caption{Measured spectra before and after the event selection is applied.
The energy scale is normalized to a calibration point of 60 keV electron recoil.}
\end{center}
\end{figure}

The WIMP observation was carried out from March to May 2005.
Measured spectra are shown in Fig \ref{spectra}.  The energy scale of 
the spectra is defined to be proportional to the number of photoelectrons.
In the energy region below 10 k.e.e., background events due to Cherenkov 
photons dominate the event rates. 
The Cherenkov events were produced in the light guides by Compton
electrons caused by background $\gamma$-rays.
We used a pulse shape discrimination (PSD) technique to eliminate these events.
Cherenkov photons are observed as fast pulses ($<$ 10 ns) while 
scintillation photons as slow pulses ($\sim$ 1 $\mu$s). 
Therefore, we used the ratio of the integral charge of the partial 
pulse shape period (0--30 ns) to that of the total pulse shape period 
(0--10 $\mu$s) 
as a discrimination parameter in the off-line analysis. 
To set the discrimination cut, 
we compared low energy Compton events which cannot 
produce Cherenkov photons (obtained using 122 keV $\gamma$-rays 
from $^{57}$Co) and Cherenkov events (obtained using 1133 and 1333 keV 
$\gamma$-rays from $^{60}$Co).
In addition, we eliminate events in which observed charge
of the two PMTs shows high asymmetry because they are thought to be
the remaining Cherenkov events or electric noise events. 
To estimate the efficiency of the off-line event selection, the low 
energy Compton 
events were used. We calculated a fraction of the events which remained 
after the selection as a function of energy. It was 0.74 at 2 k.e.e. 
Consequently, the event selection reduced the count rates 
to less than 10 counts/k.e.e./day/kg in the energy
region between 2 and 10 k.e.e.

\section{Limits on the WIMP-nucleon interaction}
Using the measured spectrum, limits on the SD WIMP-nucleon interaction were 
derived. The limits were calculated using the same manner as described in 
Ref. [16]. To derive the upper limits, we conservatively assumed 
all events to be 
nuclear recoils caused by the WIMP scattering.
The astrophysical and nuclear parameters used to obtain the limits are listed
in Table \ref{parameter}.
Only the contribution of $^{19}$F was considered while $^{40}$Ca had
negligible contribution to the SD interaction because $^{40}$Ca consisted 
of even protons and even neutrons. The scintillation efficiency $f_{\rm q}$
of the $^{19}$F recoil in a CaF$_2$(Eu) was measured by other experiments.
Although all of these experiments show that the scintillation efficiency 
tends to increase in lower energy region, we choose conservatively 
a constant value $f_{\rm q}$ = 0.11.
As was done in Ref. [17], $^{19}$F recoil energy is calculated with 
$f_{\rm q}$ and the interpolated energy scale between 0 and 60 k.e.e.: 
\begin{equation}
E_{\rm N} = \frac{L}{L_{\rm 60keV}}\frac{\rm 60[keV]}{f_{\rm q}}
\end{equation}
where $E_{\rm N}$ is the $^{19}$F recoil energy and $L$ is the scintillation 
yield of an event.

For the SD interaction, the WIMP-nucleus cross section is written as
\begin{equation}
\sigma^{\rm SD}_{\rm \chi-N} = \frac{32G^2_{\rm F}\mu^2_{\rm \chi-N}}{\pi}
(a_{\rm p}\langle S_{\rm p(N)}\rangle+a_{\rm n}\langle S_{\rm n(N)}\rangle)^2
\frac{J+1}{J},
\label{sigma_SD}
\end{equation}
where $G_{\rm F}$ is the Fermi coupling constant, $\mu_{\rm \chi-N}$ is the 
WIMP-nucleus reduced mass,
$\langle S_{\rm p(N)}\rangle$ and $\langle S_{\rm n(N)}\rangle$ are the 
expectation values of the proton and neutron spins within the nucleus and $J$
is the total nuclear spin [18]. 

The upper limit of the WIMP-nucleus scattering cross section 
$\sigma^{\rm SD}_{\rm lim \chi-N}$ is obtained by the experiment. 
From Eq. (\ref{sigma_SD}), limits in the 
$a_{\rm p}$--$a_{\rm n}$ plane is written with 
$\sigma^{\rm SD}_{\rm lim \chi-N}$:
\begin{equation}
(a_{\rm p}\langle S_{\rm p(N)}\rangle+a_{\rm n}\langle S_{\rm n(N)}\rangle)^2
\frac{J+1}{J} < \frac{\pi\sigma^{\rm SD}_{\rm lim \chi-N}}
{32G^2_{\rm F}\mu^2_{\rm \chi-N}}.
\label{ap-an eq}
\end{equation}

\begin{figure}[thb]
\begin{center}
\begin{minipage}[b]{10cm}
\includegraphics[width=10cm]{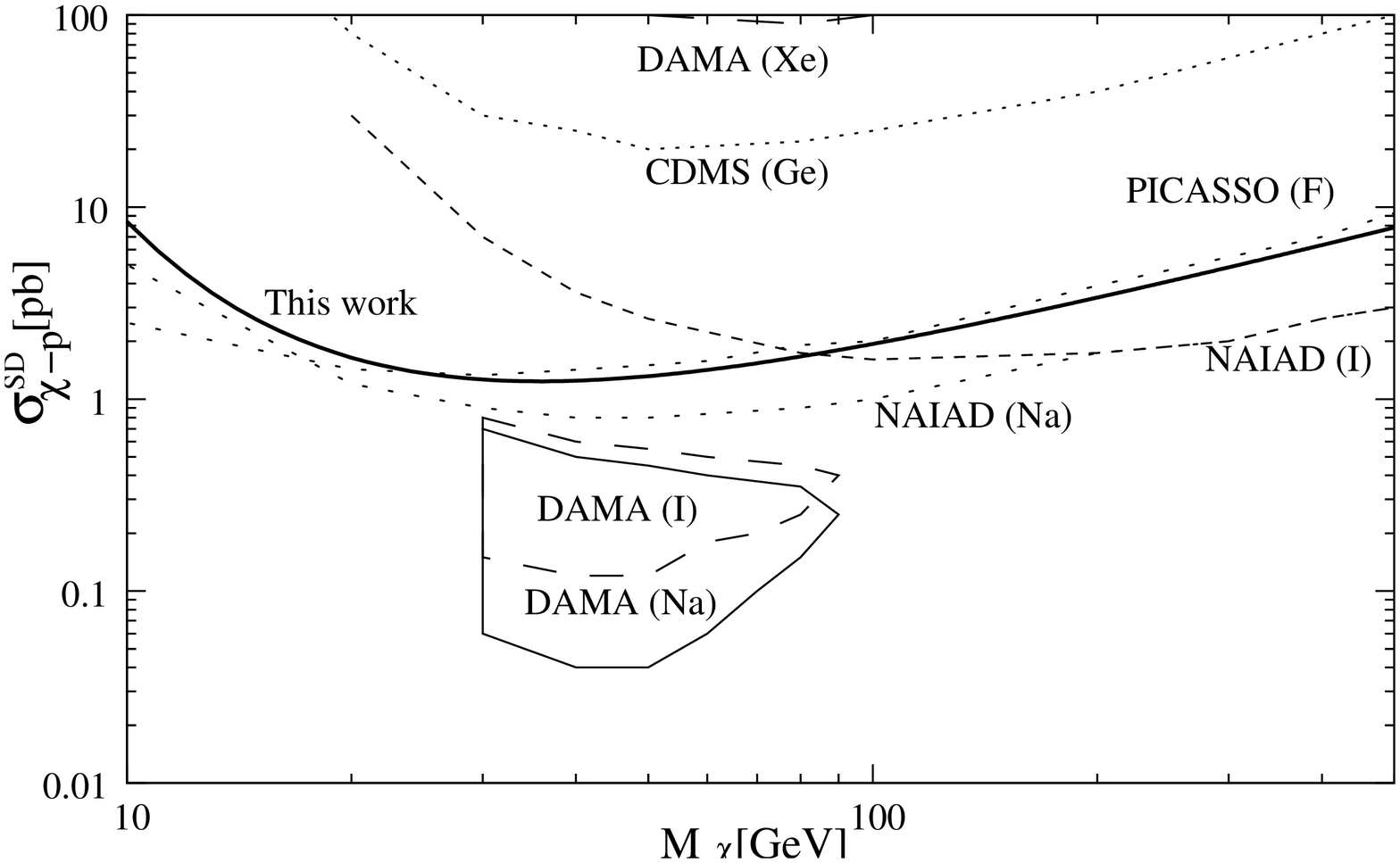}
\end{minipage}
\begin{minipage}[b]{10cm}
\includegraphics[width=10cm]{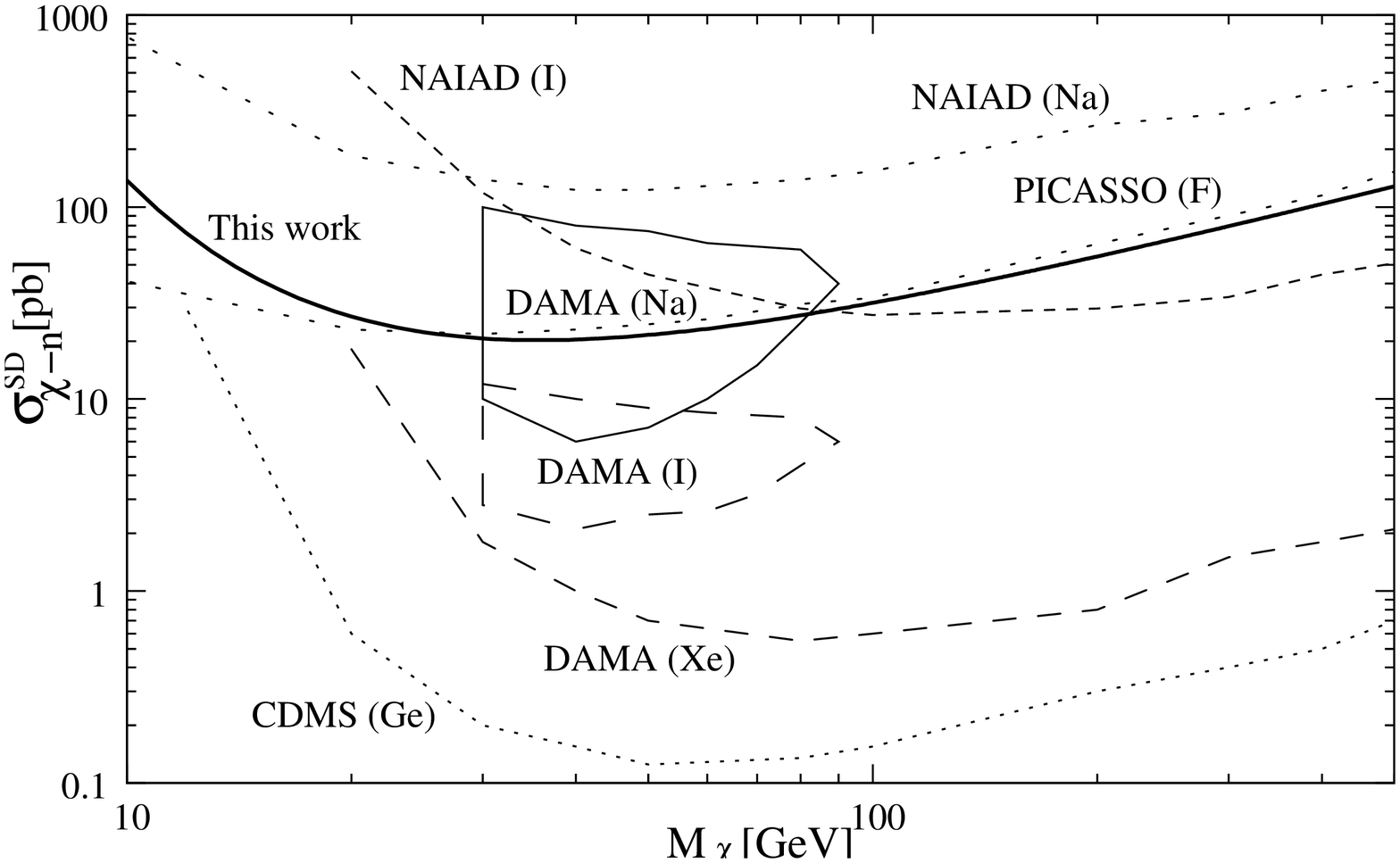}
\end{minipage}
\caption{Recent limits on the SD WIMP-proton cross section and WIMP-neutron
cross section. Results of NAIAD[20], 
PICASSO[12], CDMS[21], DAMA NaI[22], DAMA 
Xe[23] and our experiments are shown.}
\label{sigma pn}
\end{center}
\end{figure}

As in Ref. [19], limits on a single nucleon interaction are generally
calculated:
\begin{eqnarray}
\sigma^{\rm SD}_{\rm lim\chi-{\rm p(N)}} & = &
\sigma^{\rm SD}_{\rm lim\chi-{\rm N}}
\frac{\mu^{2}_{\chi - {\rm p}}}{\mu^{2}_{\chi - {\rm N}}}
\frac{{\langle S_{\rm p}\rangle}^2}{{\langle S_{\rm p(N)}\rangle}^2}
\frac{J+1}{J}, \nonumber \\
\sigma^{\rm SD}_{\rm lim\chi-{\rm n(N)}} & = &
\sigma^{\rm SD}_{\rm lim\chi-{\rm N}}
\frac{\mu^{2}_{\chi - {\rm n}}}{\mu^{2}_{\chi - {\rm N}}}
\frac{{\langle S_{\rm n}\rangle}^2}{{\langle S_{\rm n(N)}\rangle}^2}
 \frac{J+1}{J},
\end{eqnarray}
where ${\langle S_{\rm p}\rangle} = {\langle S_{\rm n}\rangle} = 
\frac{1}{2}$,
$\sigma^{\rm SD}_{\rm lim\chi-{\rm p(N)}}$ and 
$\sigma^{\rm SD}_{\rm lim\chi-{\rm n(N)}}$ are the WIMP-proton and 
WIMP-neutron scattering cross section limits when 
$a_{\rm n}\langle S_{\rm n(N)}\rangle = 0$ and  
$a_{\rm p}\langle S_{\rm p(N)}\rangle = 0$ in Eq. (\ref{ap-an eq}), 
respectively. Figure \ref{sigma pn} shows 
$\sigma^{\rm SD}_{{\rm lim}\chi-{\rm p}({\rm N})}$ and
$\sigma^{\rm SD}_{{\rm lim}\chi-{\rm n}({\rm N})}$ derived from our 
experiment as a function of WIMP mass $M_{\chi}$.
Results of other experiments are also shown in the figure. 

\begin{figure}[thb]
\begin{center}
\begin{minipage}[b]{10cm}
\includegraphics[width=10cm]{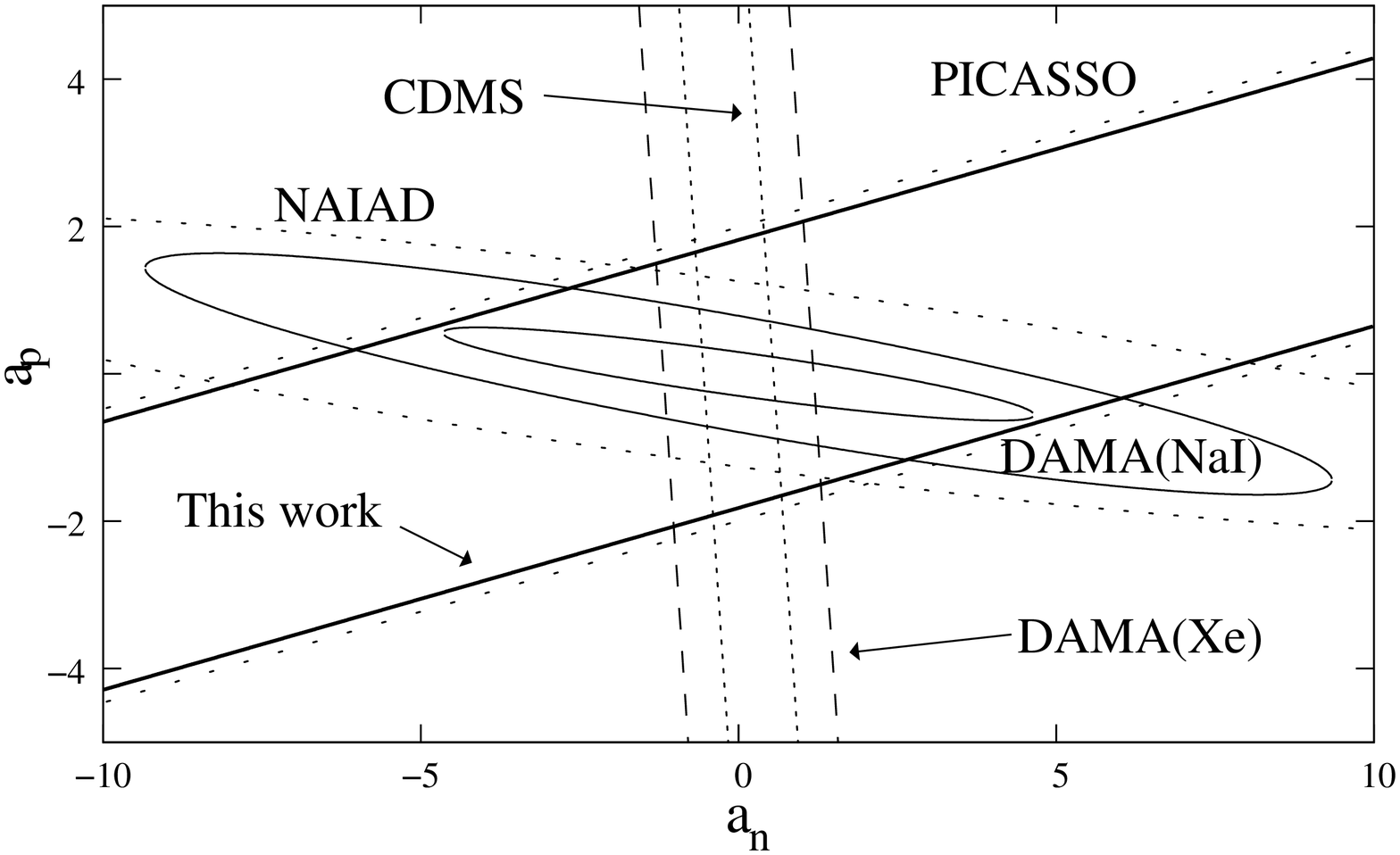}
\end{minipage}
\begin{minipage}[b]{10cm}
\includegraphics[width=10cm]{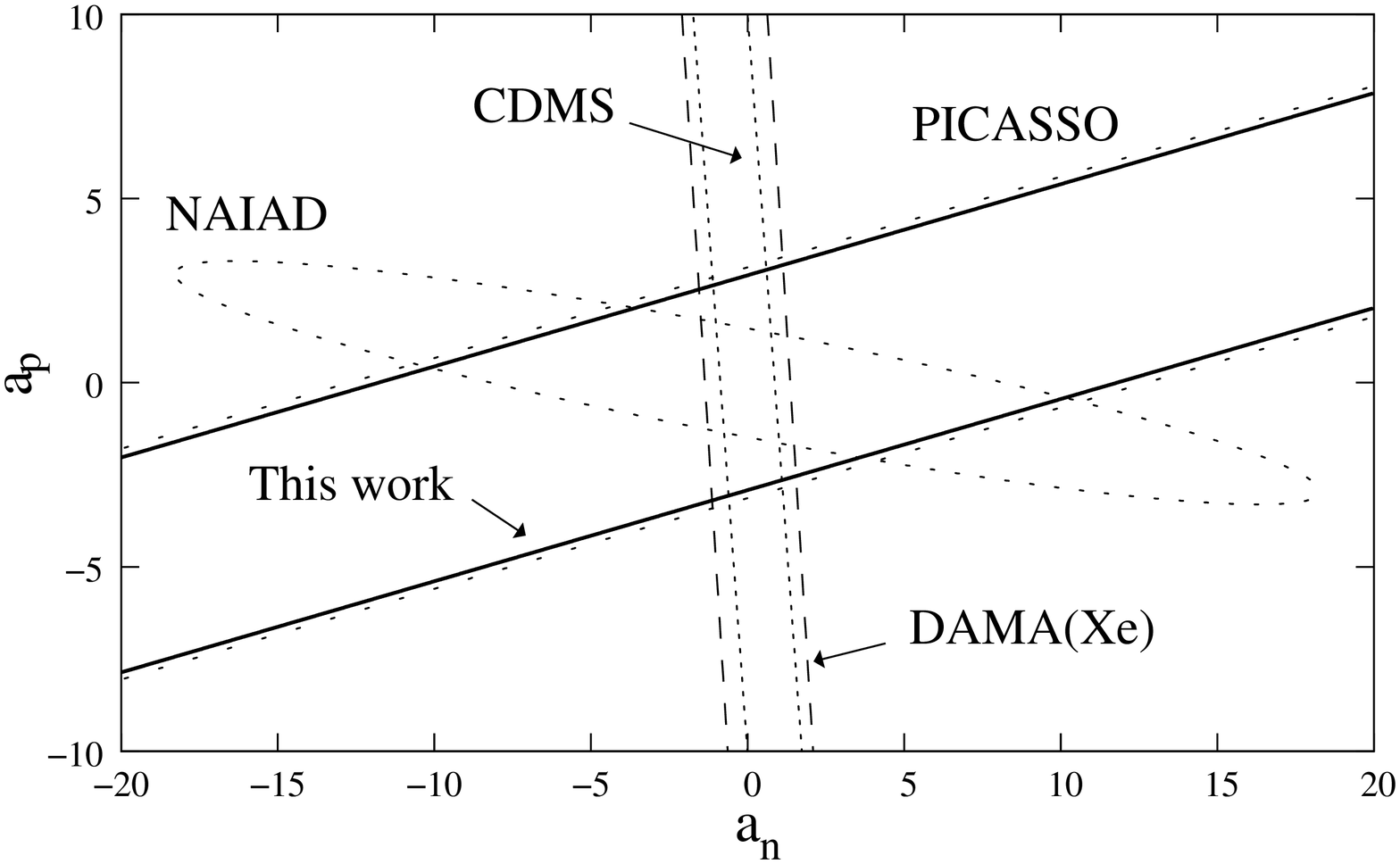}
\end{minipage}
\caption{Limits in the $a_{\rm p}$--$a_{\rm n}$ plane for $M_{\chi}$
 = 50 GeV (upper) and 200 GeV (lower). The region between two solid lines is 
allowed by this experiment. Results of NAIAD[20], 
PICASSO[12], CDMS[21], DAMA NaI[22] and DAMA 
Xe[23] experiments are also shown.}
\label{ap-an fig}
\end{center}
\end{figure}

The limits in the $a_{\rm p}$--$a_{\rm n}$ plane for WIMP mass 
$M_{\chi}$ = 50 and 200 GeV are shown in 
Fig \ref{ap-an fig}. The limits of other experiments
are also shown. The outside of the two solid lines is excluded by our 
experiment.
The region between the two ellipses is allowed by the annual modulation 
observation of the DAMA experiment [24]. 
Our results exclude a part of the DAMA allowed 
region by means of odd-proton target. Experiments with odd-neutron 
targets such as the CDMS [21] also set similar limits. 
Our results are comparable to the recent results of PICASSO [12]
experiment which uses SDDs of fluorocarbon, C$_4$F$_{10}$. 

\section{Conclusion}
The WIMP search experiment using 310g of CaF$_2$(Eu) was carried out.
The highly pure copper shield and the quartz light guides were used to
eliminate $\gamma$-rays from the outer materials. Pulse shape 
discrimination effectively eliminated Cherenkov events, which dominated
the count rates below 10 k.e.e.
As a result, count rates were lower than 10 counts/k.e.e./day/kg between 2
and 10 k.e.e.
We obtained the limits on the WIMP-nucleon spin dependent interaction in
terms of the WIMP-nucleon coupling coefficients $a_{\rm p}$ and $a_{\rm n}$.
Our results excluded a part of the parameter region allowed by the annual
modulation measurement by the DAMA NaI experiment. 

\section*{Acknowledgements}
We would like to acknowledge the help of the staff of Kamioka Observatory,
Institute for Cosmic Ray Research, University of Tokyo in performing this 
experiment. We want to express our thanks to XMASS collaboration for
allowing us to obtain the low radioactive PMT R8778 based on their 
development. We also want to express our thanks to Prof. T. Kishimoto of
CANDLES collaboration for allowing us to obtain the low background 
CaF$_2$(Eu) crystal based on the low background CaF$_2$(pure) they
developed. This research is supported by Research Center for the Early
Universe, School of Science, University of Tokyo.

\clearpage

\begin{table}
\begin{center}
\begin{tabular}[htb]{lc} \hline
Astrophysical parameters \\
Dark matter density & 0.3 GeV/cm$^3$\\
Velocity distribution & Maxwellian\\
Velocity dispersion & 220 km/s\\
Solar system velocity & 232 km/s\\
\hline
Nuclear parameters of $^{19}$F &  \\
Total spin & $\frac{1}{2}$\\
Spin expectation value of protons & $0.441$\\
Spin expectation value of neutrons & $-0.109$\\
\hline
\end{tabular}
\caption{Astrophysical and nuclear parameters used to obtain the limits.}
\label{parameter}
\end{center}
\end{table}

\end{document}